\documentclass[aps,prb,twocolumn,showpacs,superscriptaddress]{revtex4-1}

\usepackage{graphicx}
\usepackage{dcolumn}
\usepackage{bm}
\usepackage{amsmath} 
\usepackage[dvips]{color}

\newcommand{\EqLabel}[1]{\label{#1}} 
\newcommand{\mb}[1]{\mathbf{#1}}

\begin{document}

\title{A perturbational study of the lifetime of a Holstein polaron in
  the presence of weak disorder} 

\author{Hadi Ebrahimnejad} 
\affiliation{Department of Physics and Astronomy, University of
  British Columbia, Vancouver, BC, Canada, V6T 1Z1}

\author{Mona Berciu} 
\affiliation{Department of Physics and Astronomy, University of
  British Columbia, Vancouver, BC, Canada, V6T 1Z1} 
\affiliation{Quantum Matter Institute, University of British Columbia,
Vancouver, BC, Canada, V6T 1Z4}

\date{\today}

\begin{abstract}
Using the momentum average (MA) approximation, we find an analytical
expression for the disorder-averaged Green's 
function of a Holstein polaron in a three-dimensional simple cubic
lattice with random on-site energies. The on-site disorder is assumed
to be weak compared to the kinetic energy of the polaron, and is
treated perturbationally. Within this scheme, the 
states at the bottom of the polaron band are found to have an infinite
lifetime, signaling a failure of perturbation theory at these
energies. The higher-energy polaron states have a finite
lifetime. We study this lifetime and the disorder-induced
energy shift of these eigenstates for various strengths of disorder and
electron-phonon coupling. We compare our findings to the predictions
of  Fermi's golden rule and the average T-matrix method, and find a
significant quantitative discrepancy at strong electron-phonon
coupling, where the polaron lifetime is much shorter than Fermi's
golden rule prediction.  We
attribute this to the renormalization of the on-site potential by the
electron-phonon coupling.
\end{abstract}

\maketitle

\section{introduction}

Studying the behavior of solid state systems under the simultaneous
action of disorder and interactions is a significant challenge in
condensed matter physics. Strong correlations in interacting systems
often give rise to sharp quasiparticles. Scattering of
such quasiparticles from weak disorder should just limit their
lifetime. For strong disorder and no interactions, it is well
understood\cite{Anderson} that constructive interference of the
backscattered waves can localize single particles such that they lose
their itinerancy. If interactions are turned on, there is no consensus
about the effect of disorder on the quasiparticles of interacting
systems. 

For example, consider polarons, which are the focus of this work, and
which are quasiparticles comprising a charge carrier plus a cloud of
phonons describing the lattice distortions in the vicinity of the
charge. If one views the polaron as a  particle with a
renormalized mass, large disorder should result in Anderson
localization. However, phonon-assisted hopping of 
carriers between localized states is well established as a conduction
process in lightly doped semiconductors.\cite{Miller} This suggests
that, for suitable electron-phonon couplings, polarons may still be
itinerant in a disorder potential that would localize particles with
the same effective mass. A possible explanation for this was offered
recently in Refs. \onlinecite{MonaEPL} and \onlinecite{Hadi}, where
the momentum average (MA) approximation was used to
show that the electron-phonon coupling renormalizes the disorder
potential in a strongly energy-dependent manner, so that the effective
disorder seen by polarons can be drastically different from the bare
disorder.
                 
References \onlinecite{MonaEPL} and \onlinecite{Hadi} focused on the
effect of a single impurity potential on the polaron, although, as
explained there, MA
generalizes straightforwardly to other types of on-site potentials,
including disordered ones. As such, it could be used to study numerically the
effect of disorder, by generating results for various
disorder realizations and analyzing their statistics. Such results
have already been obtained for the Holstein\cite{Holstein} polaron
using the statistical dynamic mean-field theory (sDMFT).\cite{F1,F2}
Given the quite different underlying approximations, it might be
useful to check whether MA leads to similar results.

A different approach, valid for weak disorder, is to use perturbation
theory and perform the disorder average analytically, similarly to the
Born approximation widely employed for charge carriers in the absence
of electron-phonon coupling.\cite{ADM} 
Because localization cannot be described within a perturbational
calculation, the polaron eigenstates remain extended and
self-averaging over all disorder realizations is appropriate.

Here we follow the latter approach and calculate, using MA and for
weak disorder, the disorder-averaged Green's function of the Holstein
polaron and the resulting polaron lifetime and energy shift. MA is an
accurate analytical method originally developed for calculating the
Green's function of Holstein polaron in clean systems,\cite{MA0} and
later extended to other types of coupling.\cite{glen,dominic} MA is
nonperturbative in the electron-phonon coupling as it sums all
diagrams in the self-energy expansion, up to exponentially small terms
that are neglected. It also has a variational interpretation, in terms
of the allowed structure of the polaron cloud.\cite{B,MAh} MA can also
be systematically improved by increasing this variational
space,\cite{MAh} giving rise to the MA$^{(0)}$, MA$^{(1)}$, etc.,
flavors which become more accurate but at an increased computational
cost. Because here we focus only on the lowest-energy polaron states, which
are already accurately described at the MA$^{(0)}$ level, in the
following we restrict ourselves to this flavor and call it MA for
simplicity.

The paper is organized as follows. In Sec. II, we present the
generalization of MA to include disorder perturbationally. Sec. III
contains the results and their analysis, and Sect. IV has our
conclusions. Various computational  details are organized and
presented in several appendixes.

\section{The model and its solution}
The Hamiltonian for a single Holstein polaron in a lattice with random
on-site energies is
\begin{equation}
\label{Hamilt}
 	{\cal H} = {\cal H}_{\mathrm{d}} + {\hat V}_{\mathrm{el-ph}}={\cal
 	H}_0+{\hat V}_{\mathrm{d}} + {\hat V}_{\mathrm{el-ph}},
 \end{equation} 
where the non-interacting part of the Hamiltonian, ${\cal
H}_{\mathrm{d}}$, is divided into $\hat{V}_{\mathrm{d}} =\sum_i
\epsilon_i {c^{\dagger}_i} c_i$, describing the on-site disorder
potential experienced by the charge carrier, and
\begin{equation}
 	{\cal H}_0 = -t\sum_{\langle i,j\rangle}(c^{\dagger}_ic_j+H.c.) +
 \Omega\sum_i {b^{\dagger}_i} b_i,
 \end{equation} 
describing the kinetic energy of charge carrier plus the (optical) phonon energies
 ($\hbar=1$). The interaction part
\begin{equation}
\hat{V}_{\mathrm{el-ph}} =g\sum_i {c^{\dagger}_i} c_i (b^{\dagger}_i +
b_i)
 \end{equation} 
describes the Holstein coupling between the charge carrier and phonons. As
usual, $c_i$ and $b_i$ are annihilation operators for  carrier and 
phonon, respectively, on the simple cubic lattice of lattice constant
$a$, whose sites are indexed by $i$. The spin of the carrier is a
trivial degree of freedom in this model, and we ignore it. Throughout
this work we limit ourselves to the {\em single 
polaron limit}, i.e. there is a single charge carrier in the system.

The on-site energies, $\{\epsilon_i\}$, are taken from an uncorrelated
symmetric random distribution
\begin{equation}
\label{P}
 	{\cal P}(\{\epsilon_i\})=\Pi_i{\cal P}(\epsilon_i).
 \end{equation} 
For  Anderson-type disorder ${\cal P}(\epsilon_i)$
has the customary form:
\begin{equation}
\label{Anderson}
{\cal P}(\epsilon_i) =
  \begin{cases}
   1/(2\Delta) & \text{if } -\Delta\leq\epsilon_i\leq \Delta \\ 0 &
   \text{otherwise,}
  \end{cases}
\end{equation}
while ${\cal
P}(\epsilon_i)=x\delta(\epsilon_i-\epsilon_\mathrm{A})
+(1-x)\delta(\epsilon_i-\epsilon_\mathrm{B})$  
for a binary alloy, with $x$ being the concentration of A-type atoms and
energies shifted so that $x\epsilon_A + (1-x)\epsilon_B=0$. 

Our aim is to calculate the Green's function of this Holstein polaron
and average it analytically over all disorder configurations given by
Eq. (\ref{P}). From now on, we use an overbar to denote 
 disorder-averaged quantities. The strength of disorder,
 $\sigma\equiv\sqrt{\bar{\epsilon_i^2}}$, is taken to 
be weak compared to polaron bandwidth in the clean system, so that it
can be treated perturbationally. As a result, the
polaronic picture remains valid  but  its lifetime is expected to
become finite due to scattering from the disorder potential
$\hat{V}_\mathrm{d}$. It is precisely this disorder-induced
lifetime that interests us.

Since $\hat{V}_{\mathrm{d}}$ is weak compared to other terms in the
Hamiltonian, we treat it as a perturbation. Dividing the Hamiltonian
as ${\cal H}={\cal H}_{\mathrm {H}}+\hat{V}_{\mathrm{d}}$, where
${\cal H}_{\mathrm {H}}$ is the Hamiltonian of the Holstein polaron in the
clean lattice, we use  Dyson's identity, $ \hat{G}(\omega)=\hat{G}_{\mathrm
{H}}(\omega)+\hat{G}(\omega)\hat{V}_{\mathrm{d}}\hat{G}_{\mathrm
{H}}(\omega)$, to relate the resolvent $\hat{G}(\omega)$ of the system
with disorder, to  $\hat{G}_{\mathrm 
{H}}(\omega)$ of the clean system. 
To the second order in $\hat{V}_{\mathrm{d}}$, we find
\begin{eqnarray*}
\hat{G}(\omega)&\approx
&\hat{G}_\mathrm{H}(\omega)+\hat{G}_\mathrm{H}(\omega)\hat{V}_\mathrm{d}\hat{G}_\mathrm{H}(\omega)
\\
&&+\hat{G}_\mathrm{H}(\omega)\hat{V}_\mathrm{d}\hat{G}_\mathrm{H}(\omega)\hat{V}_\mathrm{d}\hat{G}_\mathrm{H}(\omega).\nonumber
\end{eqnarray*}

Because disorder breaks translational invariance, the eigenstates for
any individual disorder realization
are not labeled by the momentum ${\mb k}$. However, averaging over all
disorder configurations 
restores the translational invariance and makes momentum a good
quantum number again. As a result, $ \overline{\langle 0 | c_{\mb k}
  \hat{G}(\omega)c^{\dagger}_{\mb k'}|0\rangle} = \delta_{{\mb k},{\mb
    k'}} \bar{G}({\mb k},
\omega)$  and we only
need to calculate the diagonal matrix element:

\begin{multline}
\label{Dyson}
	\bar{G}({\mb k}, \omega) = G_\mathrm{H} ({\mb k}, \omega) + \sum_i
\bar{\epsilon_i} \langle 0 | c_{\mb k} \hat{G}_\mathrm{H}(\omega) c^{\dagger}_i c_i
\hat{G}_\mathrm{H}(\omega) c^{\dagger}_{\mb k} | 0\rangle \nonumber\\ +\sum_{i,j}
\overline{\epsilon_i \epsilon_j} \langle 0 | c_{\mb k} \hat{G}_{\mathrm{H}}(\omega)
c^{\dagger}_i c_i \hat{G}_\mathrm{H}(\omega) c^{\dagger}_j c_j
\hat{G}_\mathrm{H}(\omega) 
c^{\dagger}_{\mb k} | 0\rangle.
\end{multline}
Here, $G_\mathrm{H} ({\mb k}, \omega) = \langle 0 | c_{\mb k}
  \hat{G}(\omega)c^{\dagger}_{\mb k}|0\rangle$ is the polaron Green's
  function in the clean system. For completeness, its MA solution is
  briefly reviewed in 
  Appendix A.

Since $\bar{\epsilon_i} = 0$ for symmetric disorder, the first-order
contribution vanishes (a finite average can be removed trivially by an
overall shift of the energy).
Because of uncorrelated disorder,  $\overline{\epsilon_i \epsilon_j} =
\bar{\epsilon_i^2} \delta_{i,j} \equiv \sigma^2 \delta_{i,j}$, and the
  disorder-averaged Green's function becomes
\begin{multline*}
	\bar{G}({\mb k}, \omega) = G_\mathrm{H} ({\mb k}, \omega) \\ +\sigma^2\sum_{i}
 \langle 0 | c_{\mb k} \hat{G}_{\mathrm{H}}(\omega)
c^{\dagger}_i c_i \hat{G}_\mathrm{H}(\omega) c^{\dagger}_i c_i
\hat{G}_\mathrm{H}(\omega) 
c^{\dagger}_{\mb k} | 0\rangle.
\end{multline*}

The challenge is to use MA to calculate the matrix elements appearing in the
second term  to the same level of accuracy as $G_\mathrm{H} ({\mb k},
\omega)$.\cite{MAh}  

These matrix elements can be broken into products of generalized  Green's
functions by inserting identity operators, $\mathcal{I}$, between
the creation and annihilation operators. Since the MA flavor we use here
is equivalent with assuming that the phonon cloud only extends over
one site,\cite{MAh}  at this level of accuracy it suffices to
truncate
$\mathcal{I}\approx \sum_{l,n}(1/n!)b_l^{\dagger n} |0\rangle\langle
0|b_l^n$, i.e., to ignore states with phonons at two or more sites
(such states can be added systematically in higher flavors of
MA). This leads to:
\begin{multline}
	\bar{G}({\mb k}, \omega) =  G_{\mathrm{H}} ({\mb k}, \omega) + \sigma^2
\sum_{i,l,s,n,m} \langle 0 | c_{\mb k} \hat{G}_{\mathrm{H}}(\omega) c^{\dagger}_i
b_l^{\dagger n} |0\rangle \\ \times \langle 0|b_l^n c_i
\hat{G}_{\mathrm{H}}(\omega) c^{\dagger}_i b_s^{\dagger m} |0\rangle \langle 0|b_s^m
c_i \hat{G}_{\mathrm{H}}(\omega) c^{\dagger}_{\mb k} | 0\rangle/(n!m!). \nonumber
\end{multline} 
This expression involves two sets of generalized propagators, namely
$\langle 0 | c_{\mb k} \hat{G}_{\mathrm{H}}(\omega) 
c^{\dagger}_i b_l^{\dagger 
n} |0\rangle$ and $\langle 0|b_l^n c_i \hat{G}_{\mathrm{H}}(\omega) c^{\dagger}_i
b_s^{\dagger m} |0\rangle$.  We now evaluate them.

First, as detailed in Appendix B,  to the level of accuracy of MA, for
$n\ge 1$ 
the first propagator 
vanishes unless $i=l$. Therefore, we have to evaluate
$F^{(n)}_{{\mb k}i}(\omega)\equiv \langle 0 | c_{\mb k} \hat{G}_{\mathrm{H}}(\omega)
c^{\dagger}_i b_i^{\dagger n} |0\rangle$ for $n\ge 1$. Note that
$F^{(0)}_{{\mb k}i}(\omega)$ is already known: 
 $F^{(0)}_{{\mb k}i}(\omega)= \langle 0 | c_{\mb k} \hat{G}_{\mathrm{H}}(\omega)
c^{\dagger}_i |0\rangle= G_{\mathrm{H}}({\mb k}, \omega)\exp(-i{\mb k}\cdot
{\mb R}_i)/\sqrt{N}
$, since momentum is a good quantum number in the clean system. Here,
$N\rightarrow \infty$ is the number of sites in the system. 
The details of the calculation for $n\ge 1$, which is related  to that of
$G_{\mathrm{H}}({\mb k}, \omega)$, are presented in Appendix B. The final
result is:
\begin{equation}
     	F^{(n)}_{{\mb k}i}(\omega) =  \Gamma_n(\omega) 
     	F^{(0)}_{{\mb k}i}(\omega),
     	\label{Fn}
\end{equation} 
where $\Gamma_n(\omega)$ are easy to calculate products of continued
fractions, see Eq. (\ref{B1}).

Next, we calculate $W^{nm}(\omega)=\langle 0|b_i^n c_i
\hat{G}_\mathrm{H}(\omega) c^{\dagger}_i b_i^{\dagger m}
|0\rangle$ to the same level of accuracy. Note that because of the
invariance to translations in the clean system, this quantity is
independent of $i$. The detailed derivation of
these functions is
presented in Appendix C. 

With these expressions in hand, the disorder averaged Green's function
is, to second order in $\sigma$: 
\begin{eqnarray}
\label{Green0}
\bar{G}({\mb k},\omega)&\approx&G_{\mathrm{H}}({\mb k},\omega)+\sigma^2\left[
  G_{\mathrm{H}}({\mb k},\omega)\right]^2\times\nonumber\\ 
&&\sum_{n,m=0}^{\infty}\frac{\Gamma_n(\omega)W^{n,m}(\omega)\Gamma_m(\omega)}
  {n!m!}.
\end{eqnarray} 
To the same order, this identifies the sum in the previous equation as
the disorder self-energy,
\begin{equation}
\label{self}
\Sigma_{\mathrm{dis}}(\omega)=\sigma^2\sum_{n,m=0}^{\infty}\frac{\Gamma_n(\omega)
  W^{n,m}(\omega)\Gamma_m(\omega)}{n!m!}. 
\end{equation}
This is our main result. 
From a computational point of view, because the factorials in the
denominator grow rapidly with increasing index, the infinite
sums can be safely truncated at finite
values for $n$ and $m$. Cutoffs of 20 proved sufficient for all
cases we examined.

Using  $G_{\mathrm{H}}({\mb k},\omega)= 1/(\omega -\varepsilon_{\mb k} -
\Sigma_{\mathrm{MA}}(\omega) + i \eta)$, see Appendix A, we can
finally write
\begin{equation}
\label{Green}
\bar{G}({\mb k},\omega)=\frac{1}{\omega-\varepsilon_{\mb k}-\Sigma_{\mathrm{tot}}(\omega)
  +i\eta}, 
\end{equation}
where
the total self-energy is 
$\Sigma_{\mathrm{tot}}(\omega)=\Sigma_{\mathrm{MA}}(\omega)+\Sigma_{\mathrm{dis}}
(\omega)$. 
This implicit summation gives a more accurate expression for the
disorder-averaged 
Green's function than Eq. (\ref{Green0}), with which it agrees to
${\cal O}(\sigma^4)$.

\section{results} 
 
We are now prepared to study the effect of weak disorder on the
polaron lifetime and energy shift. At this level of perturbation
theory, disorder only enters through its standard deviation
$\sigma$. In the following, we assume Anderson disorder of width
$2\Delta$, for which $\sigma=\Delta/\sqrt{3}$. We will use either
$\Delta$ or $\sigma$ to characterize the disorder, as convenient, but
we emphasize that any other type of disorder that has the same
$\sigma$ would lead to the same answer within this perturbational
approximation.  To characterize the electron-phonon coupling strength,
it is convenient to use the effective coupling
$\lambda=g^2/(6t\Omega)$.

Once the Green's function is known, the energy broadening of a polaron
state of momentum ${\mb k}$, which is inversely proportional to its
lifetime, is the width of the low-energy peak in the spectral
function, $A({\mb k},\omega)=-\frac{1}{\pi}\mathrm{Im}\bar{G}({\mb
k},\omega)$. This broadening measures the rate at which the polaron
leaves that momentum state due to scattering from the impurity
potential $\hat{V}_\mathrm{d}$. In a clean system, the polaron states
are infinitely long lived, therefore the low-energy spectral weight is
a Dirac delta function (in fact, a Lorentzian of width
$\eta\rightarrow 0$). Mathematically, this is a consequence of the
fact that (in the absence of disorder) the polaron self-energy
$\Sigma_{\mathrm{MA}}(\omega)$ has a vanishing imaginary part for all
energies inside the polaron band.

Disorder-induced finite lifetime
broadens the delta functions into Lorentzians.
As a reference, we review first  the case without electron-phonon
coupling, $\lambda=0$.  The only nonzero term in
Eq. (\ref{self}) corresponds to $m=n=0$, therefore
\begin{equation}
\label{W00}
\Sigma_{\mathrm{dis}}(\omega)=\sigma^2W^{0,0}(\omega)=\sigma^2g_0(\omega),
\end{equation}  
where 
\begin{equation}
\label{g_0}
g_0(\omega)=\frac{1}{N} \sum_{{\mb k}} \frac{1}{\omega - \varepsilon_{\mb k}
  + i \eta}
\end{equation}  
 is the momentum-averaged free propagator.

The resulting spectral weight has a peak of width $\tau_{\mb k}^{-1}$
centred at energy $E_{\mb k}$, found from the pole condition
\begin{equation}
\label{zeros}
\omega-\varepsilon_{\mb k}-\Sigma_\mathrm{dis}(\omega)+i\eta=0,
\end{equation} 
where $\omega=E_{\mb k}-i\tau_{\mb k}^{-1}$ and $\eta\rightarrow 0^+$.

Because $\tau_{\mb k}^{-1}\sim \sigma^2$ is small for weak disorder, we approximate
$\Sigma_\mathrm{dis}(\omega)\approx \Sigma_\mathrm{dis}(E_{\mb k})+ {\cal
  O}(\sigma^4)$. Using this in Eq. (\ref{zeros}) gives $E_{\mb k}$ and 
$\tau_{\mb k}^{-1}$ as follows:\cite{note}
\begin{equation}
\begin{array}{l}
\displaystyle E_{\mb k}=\varepsilon_{\mb k}+\mathrm{Re}\Sigma_\mathrm{dis}(E_{\mb k})
\\ \displaystyle \tau^{-1}_{\mb k}=-\mathrm{Im}\Sigma_\mathrm{dis}(E_{\mb k}).
\end{array} 
\label{lifetime}
\end{equation}
The first expression determines the energy shift compared to the
electron energy in the clean system, $\varepsilon_{\mb k}$.  Since
$\mathrm{Re}g_0(\omega)$ is 
negative for $\omega<0$ and positive for $\omega>0$, this implies a
widening of the energy band in the presence of disorder.

Using Eq. (\ref{W00}), the inverse lifetime becomes 
$\tau_{\mb k}^{-1}=-\sigma^2\mathrm{Im}g_0(E_{\mb k}).$
However, $\mathrm{Im}g_0(E_{\mb k})$ is proportional to the total density of
states (DOS) for the clean system,
$$
\mathrm{Im}g_0(E_{\mb k})= -\frac{\pi}{N}\sum_{{\mb k}'}
\delta(E_{\mb k}-\varepsilon_{{\mb k}'})=-\pi\rho_0(E_{\mb k}),$$
so that: 
\begin{equation}
\label{tau}
\tau_{\mb k}^{-1}=\pi\sigma^2\rho_0(E_{\mb k})=\pi\sigma^2\rho_0(\varepsilon_{\mb k})+
    {\cal O}(\sigma^4). 
\end{equation}
The last equality is simply Fermi's golden rule. Since the density of
states vanishes outside the bandwidth of the clean system, this result
predicts infinite lifetime for all states with $|E_{\mb k}| \ge 6t$, and
finite lifetime for all states in between. We will return to this
point below.

When the electron-phonon coupling is turned on the analysis is
performed similarly, but now using the appropriate total
self-energy. The results are discussed next.

\begin{figure}[t]
\includegraphics[width=\columnwidth]{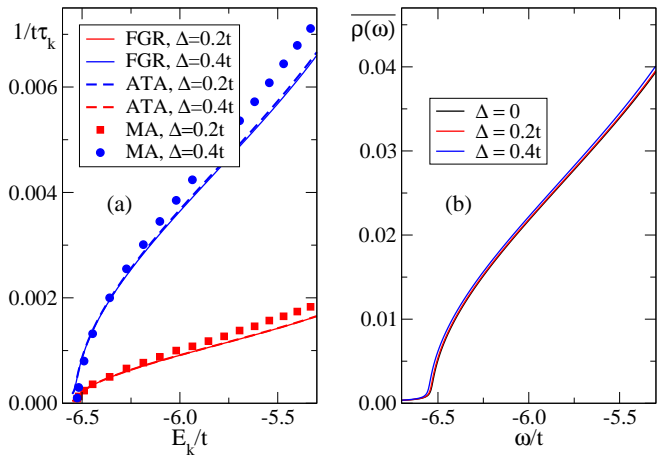}
\caption{(Color online) (a) Inverse polaron lifetime $1/\tau_{\mb k}$
  vs its peak energy $E_{\mb k}$, and
  (b) average DOS $\overline{\rho(\omega)}$ vs $\omega$, for a
  weak electron-phonon coupling and two values of the disorder
  $\Delta$. The solid and dashed lines 
  are the corresponding Fermi golden rule (FGR) and ATA results,
  respectively (see text for more details). Other parameters are
  $\Omega=t, \eta/t=10^{-2}$ in (a) and $\eta/t=5\times 10^{-3}$ in (b).}
\label{fig1}
\end{figure}

We first consider weak electron-phonon coupling, $\lambda=0.5$.  In
Fig. \ref{fig1}(a) we plot the polaron inverse lifetime for states in
the polaron band, for two different values of the disorder strength,
$\Delta=0.2t$ and $0.4t$ (squares and circles, respectively).  These
values are extracted from Lorentzian fits of the lowest peak in
$A({\mb k},\omega)$, using Eqs. (\ref{lifetime}). The broadening $\eta$ was
decreased until $E_{\mb k}$ and $\tau_{\mb k}$ converged to values independent of
it.

For this small $\lambda$, the MA ground-state energy of the
polaron in the clean system is $E_{P,GS} = - 6.534t$. The weak
disorder does not shift the eigenstates significantly. In fact, as shown in
Fig. \ref{fig1}(b), the average density of states in the disordered
system
$\overline{\rho(\omega)}=-\frac{1}{\pi}\mathrm{Im}\sum_{\mb k}\bar{G}({\mb k},\omega)$
is nearly identical to that of the clean system, although the
band becomes slightly broader with increasing $\Delta$.  The inverse
lifetime vanishes below the clean system bandedge, $E_{P,GS}$, and
above it increases
like $\sqrt{E_{\mb k}-E_{P,GS}}$, which is the expected clean system DOS at
the bottom of the band. This is very similar to the $\lambda=0$
results, except for the renormalization of the DOS by the
electron-phonon interactions. Indeed, if we think of the polaron as a
simple quasiparticle whose density of states is $\rho(\omega)$
[renormalized from $\rho_0(\omega)$ for a free electron], the inverse
lifetimes we find at the bottom of the polaron band are in good
agreement with those predicted by Fermi's golden rule (FGR), i.e., with
$\pi
\sigma^2\rho(E_{\mb k})$ (see full lines).

While the agreement between the two is good near the bottom of the band, it
becomes systematically worse at higher energies. To verify that this
amount of disorder is still sufficiently small so that the
disagreement is not due to using perturbational results outside their
validity range, we also show average T-matrix (ATA) results (dashed
lines). ATA is a simple way to treat disorder beyond the lowest order
in perturbation theory, for a system with $\lambda=0$. We briefly
discuss it in Appendix D, as well as how we extended it to finite
$\lambda$.  ATA converges to $\pi\sigma^2\rho(\omega)$ in the limit of
small $\sigma$, therefore the agreement between FGR and ATA confirms that
the contribution of higher order terms in $\sigma$ is indeed
negligible. The disagreement with MA at higher energies is, therefore,
not an artifact of 
using perturbation theory. 

The meaning of this disagreement at higher energies should, however,
be treated with some 
caution. It is well known\cite{B} that this flavor of MA fails to
reproduce the correct polaron+one phonon continuum, which should start
at $E_{P,GS}+\Omega$ (this problem is fixed by MA$^{(1)}$ and higher
flavors). One consequence is that MA overestimates the bandwidth of
the polaron at weak couplings. Indeed, in Fig. \ref{fig1}(a) we see that
the polaron band extends well past $E_{P,GS}+\Omega$. In other words, we know
that at these higher energies MA is not accurate enough, so the
results shown in Fig. \ref{fig1} should only be trusted close to the
bottom of the band, where the agreement is good.

\begin{figure}[t]
\includegraphics[width=\columnwidth]{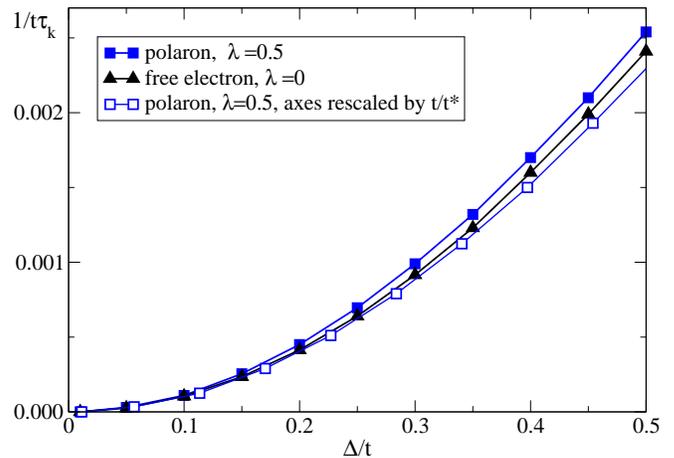}
\caption{(Color online)  Inverse lifetime of the polaron of momentum
  ${\mb k}= (\pi/8,0,0)$ and $\lambda=0.5$ vs  the strength of
  disorder, $\Delta/t$ (full squares). The inverse lifetime
  for a free electron with the same momentum is shown by triangles. 
  Empty squares show $1/(t^*\tau_{\mb k})$ vs $\Delta/t^*$ for the polaron
  (for this $\lambda$, $t^*=0.881t$). Other parameters are $\Omega=t,
\eta/t=10^{-2}$.}
\label{fig2}
\end{figure}

The monotonic decrease of the polaron's inverse lifetime with
increasing disorder strength is shown in Fig. \ref{fig2}, for a
polaron with momentum ${\mb k}=(\pi/8,0,0)$ and $\lambda=0.5$ (full
squares). For comparison, also shown is the corresponding lifetime of
a bare electron ($\lambda=0$, triangles) with the same momentum. Both
curves show the expected $\propto\sigma^2$ increase predicted by
Fermi's golden rule, but the polaron lifetime is somewhat shorter. The
most likely reason for this is the renormalization of the polaron mass
by interactions. Indeed, if instead we plot $1/(t^*\tau_{\mb k})$ vs
$\Delta/t^*$ for the polaron (empty squares), the results are much closer
to those of the free electron, especially for small values of the disorder.

The conclusion, thus far, is that Fermi's golden rule agrees well with
our results at energies where this flavor of MA can be trusted. In
other words, at weak electron-phonon coupling, the effect of 
disorder can be quantitatively understood  if we think of the polaron
as a simple 
particle with a renormalized mass (or DOS), and use Fermi's golden rule. 

\begin{figure}[t]
\includegraphics[width=\columnwidth]{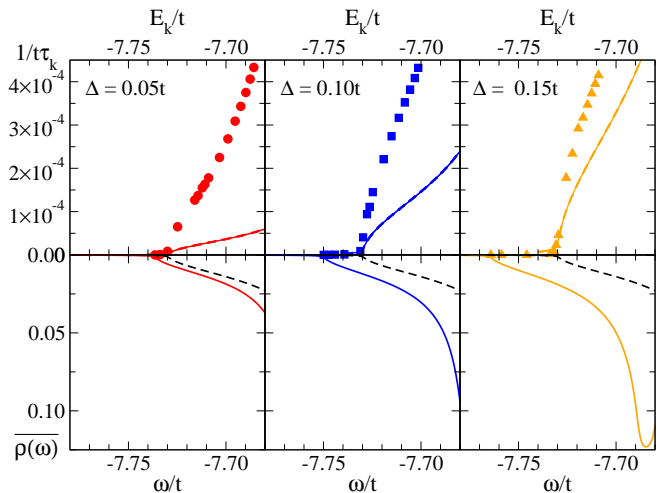}
\caption{(Color online). Top panels: $1/(t\tau_{\mb k})$ vs $E_{\mb
    k}/t$ for three levels of disorder: $\Delta/t=0.05, 0.1$ and 
  $0.15$. The symbols shows the MA result for a strong coupling
  $\lambda=1.2$, while the full and dashed 
  lines show Fermi's golden rule and the ATA predictions,
  respectively. Bottom panels: The average DOS
  $\overline{\rho(\omega)}$ for that $\Delta$ (full line) and the DOS
  in the clean system, $\rho(\omega)$ (dashed line)
  vs $\omega$. Parameters are $\Omega=t,
  \eta=10^{-3}t$.}
\label{fig3}
\end{figure}

We now check whether this also holds true at strong electron-phonon
coupling, for $\lambda=1.2$, where a robust small polaron appears in
the clean system. The top panels in Fig. \ref{fig3} show the polaron inverse
lifetime vs. its energy for three levels of disorder (symbols), as well as
the Fermi golden rule (full lines) and the ATA  (dashed
lines) predictions. The latter two are indistinguishable, confirming that these levels of disorder are indeed perturbationally small. The bottom
panels show the average density of states in the corresponding
disordered systems, $\overline{\rho(\omega)}$ (full lines). For
  comparison, the polaron DOS in the absence of disorder, $\rho(\omega)$, is
  also shown (dashed line). 

\begin{figure}[t]
\includegraphics[width=\columnwidth]{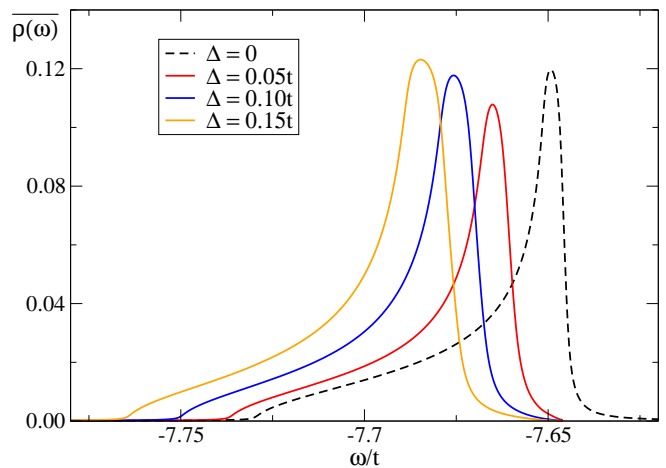}
\caption{(Color online) The same average DOS vs $\omega$ displayed in the lower panels of
  Fig. \ref{fig3}, but now shown for the entire polaron band.}
\label{fig4}
\end{figure}

Let us consider the DOS, first. As in the other cases, we see that with
increasing disorder, the bandedge shifts down to lower energies. However, the
effect is quantitatively much more significant here than at weaker
couplings because the polaron band is much narrower. This is seen in
Fig. \ref{fig4}, where we show the same densities of states but over
the full polaron band. One surprise is that the entire polaron band
moves to lower energies with increasing disorder. This is different
from what happens for a simple particle, where the band broadens
symmetrically on both sides. The different behavior at the upper edge
is likely due to the difference in their spectra. While for a simple
particle its band is the only feature in its spectrum, the spectrum of
the polaron is quite complicated, with many other  features,
such as a band associated with the second bound state, the polaron+one-phonon continuum, etc.,  lying above the polaron band.\cite{MAh} With increasing
disorder all these features should move toward lower energies. Level
repulsion from higher-energy states would explain why the upper
edge of the polaron band moves to lower energies.

For the inverse lifetime we see that, as in the other cases, it
vanishes for states with energy below the bandedge of the clean system,
$E_{\bf k} < E_{P,GS} \approx -7.73t$. Because the shift of the
disorder-averaged  DOS is now 
significant, this means that, for a quite large energy range at the
bottom of the band, the polaron has an infinite lifetime despite the
presence of disorder. We emphasize that this is qualitatively similar
to the result for a simple particle at the bottom of its band; the
effect is simply quantitatively more pronounced here. The meaning of
this (un-physical) infinite lifetime for these low-energy states is
discussed in the 
conclusions; briefly, we believe that it signals a failure of the
perturbation theory at these energies. These low-energy states are
most susceptible to localization, so the perturbational calculation and
its predictions are suspect here.

For higher-energy polaron states with  $E_{\bf k} > E_{P,GS}$,
the lifetime in the 
presence of disorder becomes finite, as expected. However, here the MA
results disagrees quantitatively with the FGR and ATA results at all
energies. The latter two are nearly indistinguishable, suggesting
again that these levels of disorder are small enough that perturbation
theory should be valid. The disagreement cannot be blamed on MA in
this case; at such strong couplings and correspondingly low energies,
MA is extremely accurate for the entire polaron band.\cite{MA0, MAh}
The disagreement is, therefore, meaningful.

Its origin can be quite easily traced. If we explicitly
separate the $n=m=0$ contribution in the disorder self-energy,
Eq. (\ref{self}) becomes
\begin{multline}
\Sigma_{\rm dis} (\omega) = \sigma^2  g_0(\omega -
\Sigma_{\rm MA}(\omega))\nonumber \\
 + \sigma^2 \sum_{n+m> 0}^{} \frac{\Gamma_n(\omega)
  W^{n,m}(\omega)\Gamma_m(\omega)}{n!m!}. \nonumber
\end{multline}

\begin{figure}[t]
\includegraphics[width=\columnwidth]{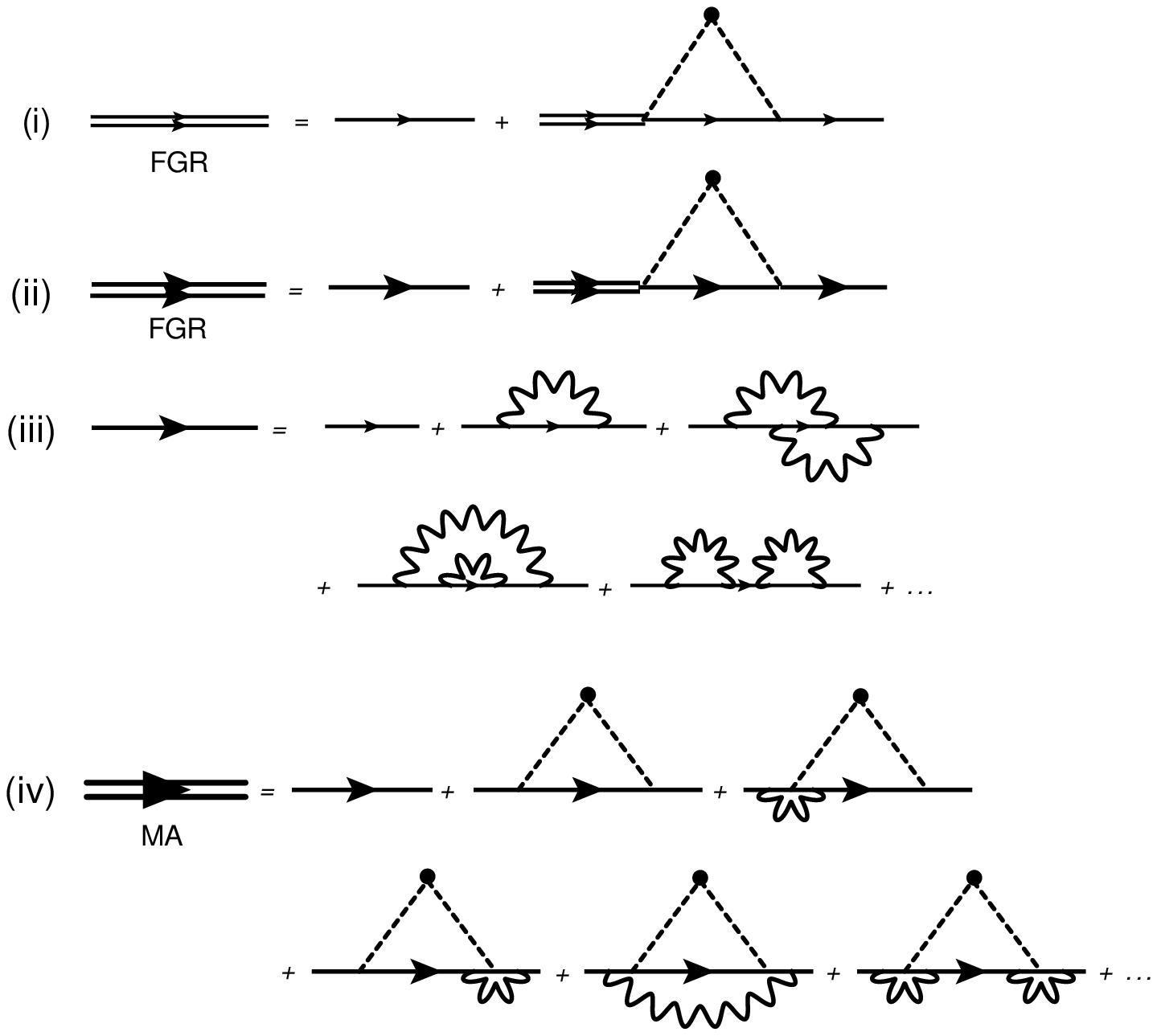}
\caption{(i) FGR approximation for the disorder-averaged Green's
  function of a carrier 
  (double thin line), in terms of that of
  the free carrier (thin line) and uncorrelated disorder (dashed
  line) in the absence of electron-phonon coupling; (ii) FGR
  approximation for the disorder-averaged Green's 
  function of a polaron (double thick line), in terms of that of the
  clean polaron (thick line); (iii) clean polaron Green's function in
  terms of free carrier (thin lines) and phonon (curly lines)
  propagators; (iv) the first few terms in the MA approximation for
  the disorder averaged Green's
  function of a polaron.
}
\label{fig5}
\end{figure} 

If the contribution of the terms with $n+m>0$ can be ignored, this
result leads to Fermi's golden rule, since $\rho(\omega) =
-{\frac{1}{\pi}} {\rm Im} g_0(\omega - \Sigma_{\rm MA}(\omega))$. The
disagreement between MA and FGR, then, comes from the contribution of
the terms with $n+m>0$. These terms cannot be ignored at strong
electron-phonon coupling. Consider, for instance, $F_{{\bf
k}i}^{(n)}(\omega)= \langle 0 | c_{\bf k} \hat{G}_\mathrm{H}(\omega)
c_i^\dagger b_i^{\dagger n}|0\rangle$, which is proportional to
$\Gamma_n(\omega)$ and therefore is responsible for its appearance in
$\Sigma_{\rm dis} (\omega)$. If we Fourier transform to real times,
$F_{{\bf k}i}^{(n)}(\tau)$ is proportional to the amplitude of
probability that if an electron is injected in the system at some
moment, at a time $\tau$ later we find the electron in the presence of
$n$ bosons, all at the same site. At strong couplings, the electron
dresses itself with a large phonon cloud to become a polaron, so the
probability of finding it with many phonons in its vicinity should be
considerable, while the probability of finding the electron without any
phonons $(n=0)$ is exponentially small. In the large $\lambda$ limit,
the terms which are expected to contribute most are those with $n
\approx g^2/\Omega^2$, i.e., values close to the average number of
phonons in the polaron cloud. In contrast, for small $\lambda$ the
phonon cloud is very fragile and, in fact, most of the time the
electron is alone (resulting in a large quasiparticle weight). This is
why, for weak coupling, keeping only the $n=m=0$ term in the sum
provides a good approximation.

These considerations are illustrated diagramatically in
Fig. \ref{fig5}. Panel (i) shows the Born approximation for the
disorder averaged Green's function of a carrier, which leads to Fermi
golden's rule expression for the lifetime, as already discussed. Panel
(ii) shows its equivalent for the disorder averaged Green's function
of the polaron; as discussed above, this is equivalent with keeping
only the $n=m=0$ term in the disorder self-energy,
Eq. (\ref{self}). Since each clean polaron propagator starts and ends
with a free carrier propagator, this approximation means that the
electron can scatter on disorder only in the absence of phonons; this
is why this approximation fails at large electron-phonon coupling,
where a large phonon cloud forms. In contrast, the full MA expression
includes diagrams such as shown in panel (iv), where phonon and
disorder lines cross. One can think of these as leading to an
effective renormalization of the disorder strength, especially since
these diagrams are very similar to those which result in the
renormalization of a single impurity potential, discussed in
Refs. \onlinecite{MonaEPL} and \onlinecite{Hadi}.

Our results in Fig. \ref{fig3} reveal that this renormalization of
the disorder (considerably) lowers the polaron lifetime. Thus, the strong
electron-phonon coupling effectively enhances the strength of the
disorder potential. Why this is so is not easy to infer from the
results for a single impurity potential, where renormalization can either
increase or decrease the bare potential, and even change its sign, and
all these behaviors are seen at different energies (retardation effects
are very significant). In any event, Fig. \ref{fig3} shows that this
renormalization is quantitatively significant.

\begin{figure}[t]
\includegraphics[width=\columnwidth]{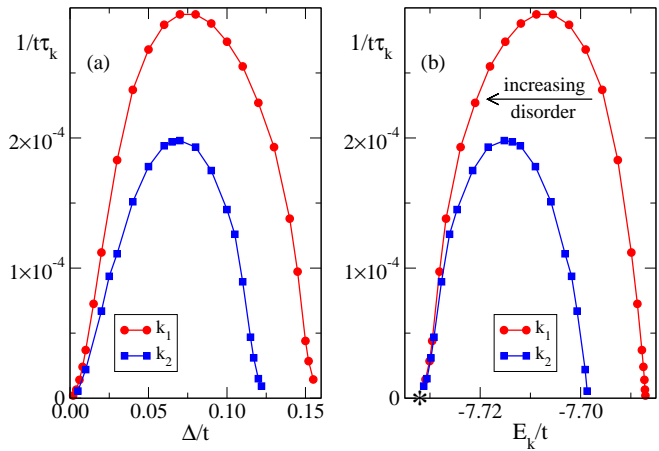}
\caption{(Color online) (a) Inverse lifetime vs disorder, and (b)
  inverse lifetime vs energy $E_{\bf k}$, as disorder is turned on, for two
  momenta ${\bf k}_1=(2\pi/9,0,0)$ and ${\bf k}_2=(\pi/6,0,0)$, for a
  polaron with 
  $\lambda=1.2, \Omega=t, \eta=10^{-3}t$. The asterisk in panel (b)
  marks the clean polaron GS energy in the clean system, $E_{P,GS}$,
  for these parameters. See text for more details.}
\label{fig6}
\end{figure}

A more detailed picture of the evolution of $\tau_{\bf k}$ and $E_{\bf k}$ with
disorder $\Delta$ at strong coupling $\lambda$ is shown in
Fig. \ref{fig6} for two 
momenta ${\bf k}_1=(2\pi/9,0,0)$ and ${\bf k}_2=(\pi/6,0,0)$, which have
 free electron energies  $\varepsilon_{{\bf k}_1}\approx-5.5t$ and
$\varepsilon_{{\bf k}_2}\approx-5.7t$, respectively. In Fig. \ref{fig6}(a),
we trace their inverse lifetimes as a function of disorder. At small
$\Delta$ both these states are well above $E_{G,PS}$, and their
scattering rates increase monotonically with $\Delta$, as one would
expect on general grounds. However, the inverse lifetimes reach a
maximum after which they begin to decrease fast and eventually
vanish. The value of $\Delta$ where they vanish is the disorder at
which that eigenstate has $E_{\mb k} = E_{P,GS}$. For larger
disorder, this eigenstate moves below the free polaron bandedge, and
its lifetime becomes infinite. 
This is more clearly shown by Fig. \ref{fig6}(b), where the inverse
lifetimes are plotted vs. the corresponding eigenenergy $E_{\bf k}$ for
these two momenta, as disorder is increased. This confirms that the scattering rates for both polaron states
vanish when their energy drops below $E_{P,GS}$, whose location is
marked by the asterisk. The value of
$\Delta$ where this happens depends on how far above $E_{P,GS}$ was
the energy $E_{\mb k}$ of this polaron, in the limit $\Delta \rightarrow 0$.

\begin{figure}[b]
\includegraphics[width=\columnwidth]{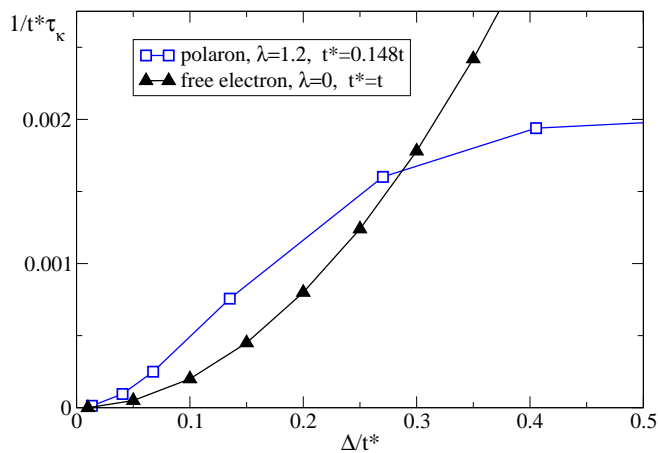}
\caption{(color online) Same data as shown in Fig. \ref{fig6}(a) but with
  rescaled axes; $1/(t^*\tau_{\mb k})$ vs $\Delta/t^*$ for momentum
  ${\mb k}_2=(2\pi/9,0,0)$ (empty squares) is compared with the free
  electron lifetime (triangles) at the same momentum, for
  $\lambda=1.2$ where $t^*=0.148t$. Other 
  parameters are $\Omega=t$ and $\eta/t=10^{-3}$.}
\label{fig7}
\end{figure}

Figure \ref{fig3} showed that using the FGR estimate, i.e., $\tau_{\mb
  k}^{-1} = \pi \sigma^2 \rho(E_{\bf k})$, is quantitatively wrong. We
can also compare the polaron lifetime, where finite, with that of a
free particle of renormalized mass, similar to the comparison in
Fig. \ref{fig2}. This is 
shown in Fig. \ref{fig7}, where we compare $1/(t^*\tau_{\mb k})$
vs $\Delta/t^*$ for the polaron, with the inverse lifetime of a free
electron with the same momentum. While roughly quadratic
dependence is observed for the polaron at small disorder, the
coefficient is quite different from that for the free
electron. At higher disorder, the disagreement is even worse. 

This shows that for intermediate and large electron-phonon coupling,
where a heavy small polaron forms, its lifetime in the presence of
disorder is not described quantitatively by
the predictions corresponding to a simple particle with renormalized
mass. The polaron has an internal structure which
manifests itself in significant corrections to Fermi's golden rule
even for weak disorder. The scattering of the electron in the presence
of its phonon cloud is quite different from that of a simple particle
of the same effective mass, but without a cloud.\cite{MonaEPL,Hadi}

\section{Summary and conclusions}

Using MA to deal with the electron-phonon coupling
and perturbation theory to deal with  the weak disorder, we derived an
expression for the disorder-averaged Green's function of the Holstein
polaron in a simple cubic lattice with random on-site energies. This
allowed us to find an analytic expression for the lowest-order
contribution from disorder to the polaron self-energy.

The disorder-averaged spectral weight was used to extract the
lifetime and energy shift of various polaron
states. For weak electron-phonon coupling, we found that the MA
results are in reasonable quantitative agreement with those predicted
by Fermi's golden rule for a free particle with an appropriately
renormalized mass.

At intermediate and larger electron-phonon coupling where a
small polaron forms, however, the MA results quantitatively disagree with
Fermi's golden rule estimate everywhere the lifetime is
finite and for all levels of disorder. The reason for this is the fact
that the scattering of the electron in the presence of 
its (robust) phonon cloud is quite different from the scattering of a
simple particle with renormalized mass. This is the same 
physics that leads to a significant renormalization of the disorder
potential seen by a polaron as compared to the bare
disorder.\cite{MonaEPL,Hadi} This demonstrates that, in
the small polaron limit, it is wrong to assume that the only
effect of the polaron cloud is to renormalize the polaron's mass.

It is important to note that this calculation is only valid for weak
disorder. It is based on perturbation theory, and in principle it can
be improved by going to higher orders along the same
lines we used to calculate the lowest-order contribution. However, one
should remember that as disorder becomes stronger, Anderson
localization will eventually occur, and that this cannot be captured
within perturbation theory. Also, once disorder is large enough to
lead to localization, the disorder-averaged Green's
function loses its meaning and usefulness. Instead, here the signature
of localization becomes manifest in the distribution of various
quantities such as the local density of states, not in their average
value. 

A surprise, at least at first sight, is the fact that this calculation
predicts an infinite lifetime for a range of energies at the
bottom of the polaron band. This interval can include a significant
fraction of the polaron states, especially at stronger electron-phonon
coupling and larger disorder. As we already mentioned, this is in fact
similar to what happens for a free particle, which also is predicted,
within this level of perturbation theory, to have an infinite lifetime
for all momenta for which $|E_{\bf k}|>6t$. The difference is only
quantitative: the energy shift for a free particle is tiny compared
with its $12t$ bandwidth, whereas for a small polaron this shift can
be comparable with its significantly narrower bandwidth even for
rather weak disorder.

A likely reason for this can be inferred from the fact that, for a
free particle, states at the band edge become localized immediately
upon introduction of disorder. In other words, we already know that
there is a finite range of energies (which, for weak disorder, falls
outside the free particle bandwidth) where treating disorder
perturbationally and calculating the disorder-averaged Green's
function is meaningless. It is then reasonable to conclude that the
states for which this perturbational scheme predicts infinite
lifetimes are, in fact, already localized. If this is correct and
generalizes to the polaron case, it suggests that, unlike for a free
particle, for a polaron localization sets in differently at the lower
vs the upper polaron bandedge. We have already speculated that this
difference may be due to the influence of the higher-energy states
that exist in the polaron spectrum. Confirmation of
these conclusions  will require a study going beyond a perturbational
treatment of disorder.

\begin{acknowledgments} This work was supported by NSERC,
CIFAR, and QMI. 
\end{acknowledgments} 

\appendix

\section{MA solution for the clean system}

Using the same notation as in the main part, the Holstein Hamiltonian
for a clean system is
 \begin{equation}
 	{\cal H}_\mathrm{H} ={\cal H}_0+\hat{V}_\mathrm{el-ph}.
 \end{equation}
 
To calculate $G_\mathrm{H}({\mb k}, \omega)= \langle 0 | c_{\mb k}
\hat{G}_\mathrm{H}(\omega) c^\dagger_{\mb k}|0\rangle$, we use Dyson's identity
$\hat{G}_\mathrm{H}(\omega) = \hat{G}_0(\omega) +
\hat{G}_\mathrm{H}(\omega)\hat{V}_\mathrm{el-ph} \hat{G}_0(\omega)$. The propagators  
for  ${\cal H}_0$ are  $\langle 0|
c_{\mb k} b_i^n
\hat{G}_0(\omega) c^\dagger_{\mb k'} b_j^{\dagger m}|0\rangle =
n!\delta_{n,m}\delta_{i,j} \delta_{\mb k, \mb k'}G_0({\mb k},\omega-n\Omega)$, where
$$
G_0({\mb k},\omega) = \frac{1}{\omega- \varepsilon_{\bf k} + i \eta}.
$$

We find
\begin{equation}
\label{Gij} 
G_\mathrm{H}({\mb k}, \omega) = G_0({\mb k}, \omega)\left[ 1 + g \sum_{i}^{}
{e^{i{\mb k}\cdot {\mb R}_i} \over \sqrt{N}}F^{(1)}_{\mb k, i}(\omega)\right],
\end{equation}
where we introduced the generalized propagators
$F^{(n)}_{\mb k, i}(\omega) = \langle 0| c_{\mb k}\hat{G}_\mathrm{H}(\omega) c_{i
}^\dagger b^{\dagger n}_i |0\rangle$. Note that $ F^{(0)}_{\mb k,
  i}(\omega)= G_\mathrm{H}({\mb k}, \omega)\exp(-i {\mb k}\cdot {\mb
  R}_i)/\sqrt{N}$.  Using the Dyson 
identity again, we find that for $n\ge 1$
\begin{multline}
\nonumber 
F^{(n)}_{\mb k, i}(\omega) = g\sum_{j\ne i}
G_0(j,i,\omega - n\Omega) \langle 0|c_{\bf k} \hat{G}_\mathrm{H}(\omega)
c_{j}^\dagger b_{j}^\dagger b_i^{\dagger n}|0\rangle  \\
\EqLabel{gg1} + g G_0(i,i,\omega-n\Omega)
\left[n F^{(n-1)}_{\mb k, i}(\omega) +F^{(n+1)}_{\mb k, i}(\omega)
\right]. 
\end{multline}
Here
$$
G_0(j,i,\omega) =  \sum_{\mb k}^{} {e^{i{\mb k}\cdot ({\mb R}_j-{\mb
	  R}_i)}\over N} G_0({\mb k}, \omega)
$$
is the free propagator in real space (the sum is over the Brillouin
zone). This propagator decays 
exponentially with the distance $|{\mb R}_j-{\mb	R}_i|$ for
energies outside the free particle continuum, $|\omega| > 6t$. Since
we are interested in energies $\omega - n \Omega \sim E_{P,GS} -
n\Omega$, where the polaron GS energy $E_{P,GS} < -6t$, all these
propagators become exponentially small for $j\ne i$. It is therefore
a reasonable first approximation to ignore $j\ne i$  terms in the EOM
written above. This is what the MA approximation does (higher flavors
include $j\ne i$ terms in a certain progression\cite{MAh}). 

Within MA, then, we have for any $n\ge 1$
\begin{equation}
\label{G-rec}
F^{(n)}_{\mb k, i}(\omega) =g g_0(\omega-n\Omega)\left[n
  F^{(n-1)}_{\mb k, i}(\omega) +F^{(n+1)}_{\mb k, i}(\omega)\right], 
\end{equation}
where $g_0(\omega) = G_0(i,i,\omega)$; see Eq. (\ref{g_0}). This
recurrence relation is solved in terms of continued fractions:\cite{MA0}
\begin{equation}
     F^{(n)}_{\mb k, i}(\omega)  = A_n(\omega) F^{(n-1)}_{\mb k, i}(\omega)
     	\label{Fnn}
\end{equation} 
for any $n\ge 1$, where
\begin{equation}
	A_n(\omega)=\frac{ngg_0(\omega-n\Omega)}{1-gg_0(\omega-n\Omega)A_{n+1}(\omega)}.
	\end{equation}
Finally, using  $ F^{(1)}_{\mb k, i}(\omega) = A_1(\omega)
F^{(0)}_{\mb k, i}(\omega)$ in  Eq. (\ref{Gij}) leads to the MA
solution: 
$$
G_\mathrm{H}({\mb k},\omega) = {1\over \omega - \varepsilon_{\mb k} -
  \Sigma_\mathrm{MA}(\omega) + i \eta}
$$
where
 \begin{equation}
\Sigma_\mathrm{MA}(\omega)=gA_1(\omega).
\end{equation}
  
\section{MA solution for $F_{{\mb k}i}^{(n)}(\omega)$}

As discussed in the text, we need to calculate the propagators:
 $\langle 0 | c_{\mb k} \hat{G}_\mathrm{H} (\omega)c^{\dagger}_i
 b^{\dagger n}_s |0\rangle$. Within the MA approximation (see above),
 this propagator is set to zero for all $i\ne s$, because it is
 proportional to $G_0(s,i,\omega-n\Omega)$. The only finite value is
 for $i=s$, in which case $\langle 0 | c_{\mb k} \hat{G}_\mathrm{H}
 (\omega)c^{\dagger}_i b^{\dagger n}_i |0\rangle = F_{{\mb
 k}i}^{(n)}(\omega)$. These propagators have already been calculated above,
 $F^{(n)}_{\mb k i}(\omega) = A_n(\omega) F^{(n-1)}_{\mb k i}(\omega)
 = \dots = \Gamma_n(\omega) F^{(0)}_{\mb k i}(\omega)$, where
\begin{equation}
\label{B1}
     \Gamma_n(\omega)\equiv A_n(\omega)A_{n-1}(\omega)\cdots A_1(\omega).
\end{equation} 

\section{MA expression for $W^{n,m}(\omega)$}
Here we calculate the remaining needed propagators, $W^{nm}(\omega)=\langle 0|b_i^n c_i
\hat{G}_\mathrm{H}(\omega) c^{\dagger}_i b_i^{\dagger m}
|0\rangle= W^{mn}(\omega)$. Because of this symmetry, we only need to
find $W^{nm}(\omega)$ for $m\le n$. 
Note that we already know  $W^{00}(\omega)=\langle 0|c_i
\hat{G}_\mathrm{H}(\omega) c^{\dagger}_i 
|0\rangle = {1\over N} \sum_{{\mb k}}^{}G_\mathrm{H}({\mb k}, \omega) =
g_0(\omega- \Sigma_\mathrm{MA}(\omega))$, and also $W^{n0}(\omega)=\langle 0|b_i^n c_i
\hat{G}_\mathrm{H}(\omega) c^{\dagger}_i
|0\rangle = \sum_{{\mb k}}^{} {e^{-i{\mb k}\cdot {\mb R}_i}\over
  \sqrt{N}} \left[F_{{\mb k}i}^{(n)}(\omega)|_{\eta \rightarrow
	-\eta}\right]^* = \Gamma_n(\omega) g_0(\omega- \Sigma_\mathrm{MA}(\omega))$.

For any $m\ge 1$, writing the EOM for $W^{nm}(\omega)$ within the MA
approximation (i.e., not allowing the electron to change its site),
leads to
\begin{eqnarray}
\label{W}\nonumber
W^{n,m}(\omega)&=&m!g_0(\omega-m\Omega)\delta_{nm}+gg_0(\omega-m\Omega)\\
&& \times[mW^{n,m-1}(\omega)+W^{n,m+1}(\omega)].
\end{eqnarray}
For $m>n$, the delta function vanishes and
Eq. (\ref{W}) is identical to  Eq. (\ref{G-rec}), hence
\begin{equation}
 W^{n,m}(\omega)=A_m(\omega)W^{n,m-1}(\omega).
\end{equation}
In particular, this gives
$W^{n,n+1}(\omega)=A_{n+1}(\omega)W^{n,n}(\omega)$.  Using
this in Eq. (\ref{W}) with $n=m$ relates $W^{n,n}(\omega)$ to
$W^{n,n-1}(\omega)$:
\begin{equation}
 W^{n,n}(\omega)=A_n(\omega)W^{n,n-1}(\omega)+\frac{(n-1)!}{g}A_n(\omega).
\end{equation}

This is taken together with the EOM for $1\le m\le n-1$
$$
W^{n,m}(\omega)=gg_0(\omega-m\Omega)\times[mW^{n,m-1}(\omega)+W^{n,m+1}(\omega)]
$$ to give a system of $n$ equations with $n$ unknowns
$W^{n,m}(\omega)$, $m=1,..,n$ [$W^{n,0}(\omega)$ is
known; see above]. This can be solved in many ways, including direct numerical
solution. A nicer approach is to use the linearity of this system of
equations to split it into two different systems, one which has only
$\frac{(n-1)!}{g}A_n(\omega)$ and one which has only
$gg_0(\omega-\Omega)W^{n,0}(\omega)$ as inhomogeneous
  parts. These can be solved analytically, to give
$$
W^{n,m}(\omega) = \Gamma_m(\omega) W^{n,0}(\omega)
+\tilde{\Gamma}_m(\omega)\frac{(n-1)!A_n(\omega)}{g[1-A_n(\omega)B_n(\omega)]},
$$
where $\tilde{\Gamma}_m(\omega) = B_{m+1}(\omega)B_{m+2}(\omega)\cdots
B_n(\omega)$ for $m< n$ while $\tilde{\Gamma}_n(\omega)=1$, and
$$
B_{m+1}(\omega) = \frac{g g_0(\omega-m\Omega)}{1- (m-1)g
  g_0(\omega-m\Omega) B_{m}(\omega)}
$$
are continued fractions ending at $B_2(\omega) = g g_0(\omega
-\Omega)$.

\section{ATA self-energy}
ATA\cite{Elliot,book} relates the disorder part of the self-energy of
a single particle to the 
disorder average of its transfer matrix through a single impurity with
on-site energy $\epsilon$:
\begin{eqnarray}
\Sigma_\mathrm{ATA}(\omega)=\frac{\bar{t}}{1+\bar{t}g_0(\omega)},
\end{eqnarray}
where $t=\epsilon/(1-\epsilon g_0(\omega))$ is the sum over all single
impurity scattering contributions.  $g_0(\omega)$ is the momentum
average of the free particle 
propagator, see Eq. (\ref{g_0}).

For  Anderson-type disorder we find
\begin{eqnarray}
\bar{t}&=&\frac{1}{2\Delta}\int_{-\Delta}^\Delta \frac{\epsilon
d\epsilon}{1-\epsilon g_0({\omega})}\nonumber\\
&=&-\frac{1}{g_0({\omega})}+\frac{1}{2\Delta
g_0^2({\omega})}\ln \frac{1+\Delta g_0({\omega})}{1-\Delta
g_0({\omega})}.\nonumber
\end{eqnarray}

Expanding to lowest order regains the
perturbational result, Eq. (\ref{W00}):
$\Sigma_\mathrm{ATA}(\omega)\approx 
\frac{\Delta^2}{3}g_0({\omega})=\sigma^2g_0({\omega}) $. Differences
between ATA and FGR show that disorder is so large  that
multiple scattering processes off the same impurity cannot be ignored anymore.

To extend ATA to the Holstein model, we note that difference between
the MA clean polaron's Green's function and that of the free electron
is the appearance of $\Sigma_\mathrm{MA}(\omega)$. This simply modifies
$g_0(\omega) \rightarrow g_0(\omega-\Sigma_\mathrm{MA}(\omega))$ in all the
above equations. As discussed above, this approximation implies that there
is no crossing between phonon lines and scattering lines, in
diagramatic terms. Our results show that this is a bad approximation
at larger electron-phonon coupling.


\begin{thebibliography}{99}
\bibitem{note} Strictly speaking, the lifetime is defined with an
  additional factor of 2.
\end{thebibliography}
\end{document}